\documentstyle[twocolumn,prb,aps]{revtex}
\begin{document}

%2
\flushbottom
%2
\twocolumn[\hsize\textwidth\columnwidth\hsize\csname 
%2
@twocolumnfalse\endcsname 

\title{Electronic Structure and Thermoelectric Prospects of Phosphide
Skutterudites} 

\author{Marco Fornari}
\address{C.S.I. George
Mason University, Fairfax, VA 22030-4444 and Naval Resarch Laboratory,
Washington D.C. 20375-5345}   
\author{David J. Singh}
\address{Complex Systems Theory Branch, Naval Research Laboratory,
Washington D.C. 20375-5345}

\date{To be published on Phys. Rev. {\bf B}}
\maketitle

\begin{abstract}
%1\setlength{\baselineskip}{0.7cm}
The prospects for high thermoelectric performance in phosphide skutterudites
are investigated based on first principles calculations. We find that 
stoichiometric $CoP_3$ differs from the corresponding arsenide and
antimonide in that it is metallic. As such the band structure must be
modified if high thermopowers are to be achieved. In analogy to the 
antimonides it is expected that this may be done by filling with
La. Calculations for $LaFe_4P_{12}$ show that a gap can in fact be
opened by La filling, but that the valence band is too light to yield
reasonable p-type thermopowers at appropriate carrier densities;
n-type $La$ filled material may be more favorable. 
\end{abstract}

%2
]
%1
\newpage

% Introduction
There has been considerable recent interest in the electronic and
thermal transport properties of skutterudites. This is driven 
primarily by the discovery of two new high performance thermoelectric
(TE) materials in this class.\cite{mss-pt,smw-s} The TE performance is
characterized by a dimensionless figure of merit $ZT=
\sigma S^2 T/k$, where $\sigma$ is the electrical conductivity,
$S$ is the thermopower and $k$ is the thermal
conductivity; $ZT$ up to 1.4 at $T=$ 600 K has been measured in skutterudites.
Much of the effort has focused on antimonides \cite{fcb-mrs}
based on the expectation of lower values of the lattice thermal
conductivity related to the heavier mass atoms as well as the
likelihood of better carrier mobilities due to the chemistry of Sb as
compared to say P. In fact, the two high $ZT$ compositions discovered
are both antimonides\cite{fbcmm-ict,cfb-ict}: $CeFe_4Sb_{12}$  and
$La(Fe,Co)_4Sb_{12}$.

The high $ZT$ values in these compounds derive from two important
features: (1) high power factors $\sigma S^2$ related to their
particular electronic 
structures, which are apparently different both between the two
compounds and from the corresponding binary, $CoSb_3$; and (2) a strong
suppression of the thermal conductivity of the binary upon
filling. This latter effect, though crucial for the TE performance, is
understood only qualitatively in terms of phonon scattering related to
rare earth vibrations. Attempts to obtain even better
performance by various alloying and substitutions on each of the three
sites have thus far been unsuccessful, although there are still many
possibilities remaining to be explored. These efforts are complicated
by the large variety of realizable modifications of these
skutterudites and the general lack of detailed understanding of their
effects on properties relevant to TE. Moreover, TE performance
typically is a strong function of the doping level, further
complicating the search.  

In the present brief report we present electronic structure
calculations for the phosphides, $CoP_3$ and $LaFe_4P_{12}$ and discuss these
in terms of the implications for TE performance and in relation to the
corresponding antimonide materials in order to elucidate trends. All
previous first principles calculations point to a particularly
important role for bands associated with the chemical bonding of the
pnictogen 4-membered rings 
in the skutterudite structure in determining transport properties - a
point that 
was emphasized early on by Jung et al. \cite{ywa-ic} based on tight binding
calculations. Calculations for $CeFe_4P_{12}$, $CeFe_4As_{12}$ and
$CeFe_4Sb_{12}$ \cite{ns-prb} have 
shown these materials to be hybridization gap semiconductors with
decreasing gaps as the lattice parameter increases and $Ce$-{\it f}
hybridization decreases down the pnictogen column. Previous
calculations for the binaries $CoSb_3$ and $CoAs_3$ reveal generally similar
electronic structures, \cite{sp-prb} but with differences that are particularly
significant in the region near the Fermi energy (E$_F$) that dominates
electronic transport. $CoSb_3$ is a narrow gap semiconductor with a
highly non-parabolic valence band dispersion, while $CoAs_3$ was found to
be a zero gap semiconductor with parabolic bands. 

Zhukov has reported
first principles band structure calculations for $CoP_3$ finding the
material to be a narrow indirect gap semiconductor
\cite{z-pss,laaz-prb}. The relatively
heavy conduction bands with their multi-valley minima would seem
initially favorable for the electronic aspect of TE performance with
n-type doping. However, the calculations were done with the linear
muffin tin orbital atomic sphere approximation method
(LMTO-ASA). Because the skutterudite crystal structure features large
voids, low site symmetries and strong covalent bonding, such
calculations are particularly difficult, and in such cases may have
band shifts of several tenths of an eV compared to more accurate
general potential calculations. Because of the small indirect gap, this is
enough to qualitatively change the picture from a transport point of
view, implying the need for a general potential investigation as
presented here. 

Our calculations were done in the framework of  density
functional theory using the general potential linearized
augmented plane wave (LAPW) method\cite{s-lapw} which does not make
any shape approximations and uses a flexible basis set including LAPW
functions and local orbital extensions\cite{s-prb} to
relax linearization errors and treat semicore-states. Valence states
were done in a scalar relativistic scheme while fully relativistic
calculations were done for core states 
in the atomic 
spheres (R$_{MT}$ = 2.1, 2.1, 1.9, 2.5 a.u. for $Co$, $Fe$, $P$ and $La$
respectively). The the basis set convergence was tested using 
$R_{min}\cdot k_{MAX}$ from 5.0 to 8.5; a value of 7.0 was found to
yield a reasonable computational effort with only a small
error with respect the highest $R_{min}\cdot k_{MAX}$ ($\Delta E_T =
0.3\ \frac{mRy}{atom}$).
We used a (4,4,4) special points grid for the Brillouin zone
integration, which we found to be 
converged.  The electronic density
of states (DOS) was based on a 35 {\bf k} points tetrahedral sampling
in the irreducible BZ . The Hedin-Lundqvist
parameterization for the  exchange-correlation LDA functional is used.

As discussed below, we find, for $CoP_3$, a globally similar band
structure to Zhukov, but with changes near E$_F$ that are large enough to
drastically change the picture in an unfavorable direction from the
point of view of TE.

As mentioned, the binary $CoSb_3$ has a
relatively high thermal conductivity and a highly non-parabolic valence band dispersion,
which is unfavorable for high  p-type TE performance. Meanwhile
$La(Fe,Co)_4Sb_{12}$ has both a strongly reduced $k$ and a band structure that
is modified in such a way as to improve the electronic properties by
shifting the valence band edge downwards due to repulsion from the La
{\it f}-resonance above the Fermi level \cite{sm-rprb}. One may
conjecture that a similar 
effect could be present in $La(Fe,Co)_4P_{12}$ as the band edge states have the
same character, and if so the question arises as to whether the
electronic properties relevant to TE may be improved.
There is also  interest in the electronic structure of filled
phosphide skutterudites because of the observation of
superconductivity in $LaFe_4 P_{12}$ with critical temperature,
T$_c$ = 4.1 K \cite{m-phy}. 

% Method
The skutterudite structure (space group $Im\overline{3}$)
consists of a simple cubic transition metal sub-lattice partially
filled by almost square pnictogen groups ($P_{4}$). Three quarters of
these sites are filled with such rings oriented in [100], [010] and [001]
directions according to the cubic symmetry and the remaining one
quarter are left empty. In filled skutterudites these remaining sites
are occupied by a rare earth ion,
which modifies thermal and electronic properties. 
Two symmetry independent parameters $u$ and $v$ determine the position of
the $P$ with respect to the metal ion in the center of the cubic
cell; they control the size and the squareness of the rings. 
We start from experimental crystal structure fixing
the position of pnictonen group with respect the transition metal to
$u_e=$ 0.1453 $a$ and $v_e=$ 0.3482 $a$ where the lattice parameter is $a$
= 0.77073 nm for $CoP_3$ and $u_e=$ 0.1504 $a$ and $v_e=$ 0.3539 $a$ where
a = 0.78316 nm for $LaFe_4 P_{12}$\cite{hand}.

% CoP3
The band structure of $CoP_3$, as given in Fig.\ref{BS1}, is metallic
due to the fact that the pseudogap near E$_F$, which is characterstic
of skutterudites, is entirely crossed by a single mostly phosphorus
{\it p} band. This is the same band as crosses the pseudogap in
$CoSb_3$ and $CoAs_3$, but in $CoP_3$ it crosses the conduction bands
above the Fermi level.
As such, $CoP_3$ is not very promising for TE
applications unless filling or other modifications alter the band
structure enough to open a gap.
The corresponding DOS and projections are shown in Fig.\ref{DOS}. 
In the most relevant region for TE
properties (near E$_F$) our LAPW band structure is quite different
from the LMTO-ASA results\cite{laaz-prb} which predict an indirect $\Gamma$-$H$
energy gap.
As already found in Ref.\onlinecite{sp-prb} for antimonides, there are
(1) a single 
degenerate band, (2) a two-fold degenerate band and (3) a three-fold
degenerate band at $\Gamma$ and above E$_F$. The first one is mostly P
{\it p}-derived while the
other two are more hybridized with higher contributions from $Co$ {\it
d}-states.  
The energy alignments of these bands at $\Gamma$ point are different
for different $Co$ derived skutterudites passing from (1)-(3)-(2) in
$CoSb_3$ to (3)-(2)-(1) in $CoAs_3$ to (2)-(1)-(3) in $CoP_3$.

What we need to ``realize'' a favorable band structure for p-type TE
properties from $CoP_3$ is to lower the single
degenerate band (1) until it crosses band (2) and arrives near to the
heavy mass bands forming the bottom of the pseudogap so that we obtain
a semiconductor with potentially high Seebeck coefficients analogous
to $La(Fe,Co)_4 Sb_{12}$.
The effect of gap opening by means of filling that provides
TE performance for antimonides relies
on the interaction between rare earth {\it f}-states or  resonances
and the crossing band. This effect is strong because the wavefunctions of
the pseudogap crossing band have {\it f}-like symmetry as discussed in
detail in Ref. \onlinecite{sm-rprb}.

% LaFe4P12
Our results on $LaFe_4 P_{12}$ show that the
crossing band is indeed pushed down, as expected, by repulsion of $La$
{\it f}-resonance
states at 
about 3.0 eV above E$_F$ (Fig.\ref{DOS} lower panel) resulting
in a small direct gap between
bands (1) and (2): $E_g = 98$ meV. However the top of the band crossed
by the Fermi level is not 
close to any lower heavy bands (Fig.\ref{BS2}) so that p-type thermopower can
not be high enough at a reasonable band filling for $La(Fe,Co)_4
P_{12}$ assuming rigid band behavior upon alloying with $Co$ as found
in $La(Fe,Co)_4 Sb_{12}$ (N.B. strong non-rigid band behavior, which
we do not expect, would also be detrimental to TE as it would indicate
strong alloy scattering and low carrier mobility).
A study of the effective masses (Tab.\ref{TAB1}) for the three bands
closest to $E_F$ 
points out also that because of the double degeneracy and the
reasonably high $m^{\star}$,  $La(Fe,Co)_4 P_{12}$
could be more interesting for n-type TE application, but only if
thermal conductivity can be strongly reduced and well-filled n-type
material with low defect concentrations and high mobility can be produced.

The importance of four membered pnictogen rings for thermoelectricity and also
for superconductivity \cite{m-phy} suggests investigation of the $A_g$
Raman active phonon frequencies\cite{lk-zaac} at 
Brillouin zone center that are associated to normal modes involving
variations of the symmetry-independent parameters ($u$,$v$) in the
skutterudites structure or, in other words, distortion of the
pnictogen rings. We obtain LDA structural parameters ($u_0=$ 0.1462 and $v_0=$ 0.3478)  near the
experimental ones and $\omega_1=$
228 cm$^{-1}$ and $\omega_2=$ 177 cm$^{-1}$ for $CoP_3$. 
Similar calculations for $LaFe_4 P_{12}$ give $u_0=$ 0.1537 and $v_0=$
0.3522 for LDA equilibrium parameters whereas $\omega_1=$ 189 cm$^{-1}$
and $\omega_2=$ 160 cm$^{-1}$. The difference presumably reflects $La$-$P$ interactions. 

% Conclusions
In summary we have presented electronic structure calculations for
$CoP_3$ and $LaFe_4 P_{12}$. These show that  while $CoP_3$ is a
metal, a gap is opened up upon filling with $La$. Nonetheless the band
structure does not allow for high p-type thermopowers with reasonable
carrier concentrations. The conduction bands are more favorable
having a degenerate heavy mass structure, though we note that n-type
filled skutterudites are difficult to prepare.

We are grateful for useful discussions with
R. S. Feigelson. This work is supported by ONR and DARPA.

\bibliographystyle{/usr/local/Revtex/prsty}

\newpage

\begin{table}%%%%%%%%%%% tab 1
\begin{tabular}{|c|c|c|c|} 
E($\Gamma$) & $m^{\star}(\Gamma - H)$ & $m^{\star}(\Gamma - P)$ &
$m^{\star}(\Gamma - N)$       \\ \hline
-0.1137 & -0.99 & -2.67 & -1.61 \\ %\cline{2-4}
        & -2.67 &       & -2.85 \\ \hline
 0.2597 & -0.93 & -1.14 & -1.12 \\ \hline
 0.9887 & -0.32 & -0.37 & -0.34 \\ \hline
 1.0697 &  3.02 &  1.24 &  1.70 \\ %\cline{2-4}
        &  0.56 &       &  0.93 \\  
\end{tabular}
\caption[TAB. 1]{Effective masses for $LaFe_4P_{12}$ along the high symmetry
directions in BZ. E($\Gamma$) denotes the band energy (in eV) at
$\Gamma$ relative to E$_F$ (see Fig.\ref{BS2}).}
\label{TAB1}
\end{table}

\begin{figure}%%%%%%%%%%% fig 1
\caption{Band structure for $CoP_3$ in an energy window centered at
E$_F$. Notice the metallic character due to band crossing at $\Gamma$.}
\label{BS1}
\end{figure}

\begin{figure}%%%%%%%%%%% fig 2
\caption{DOS for $CoP_3$ (upper panel) and for $LaFe_4P_{12}$ (lower
panel), the energy is referred to the respective Fermi level. The
total DOS (solid line) and the projections on different relevant
atomic components that are sketched: P {\it p}-components (light gray
shadow), $Co$ {\it d}-components (dashed line) and resonant La {\it
f}-components (dark gray shadow).} 
\label{DOS}
\end{figure}

\begin{figure}%%%%%%%%%%% fig 3
\caption{Band structure for $LaFe_4P_{12}$ in an energy window including 
E$_F$. Notice the small gap opened at $\Gamma$.}
\label{BS2}
\end{figure}

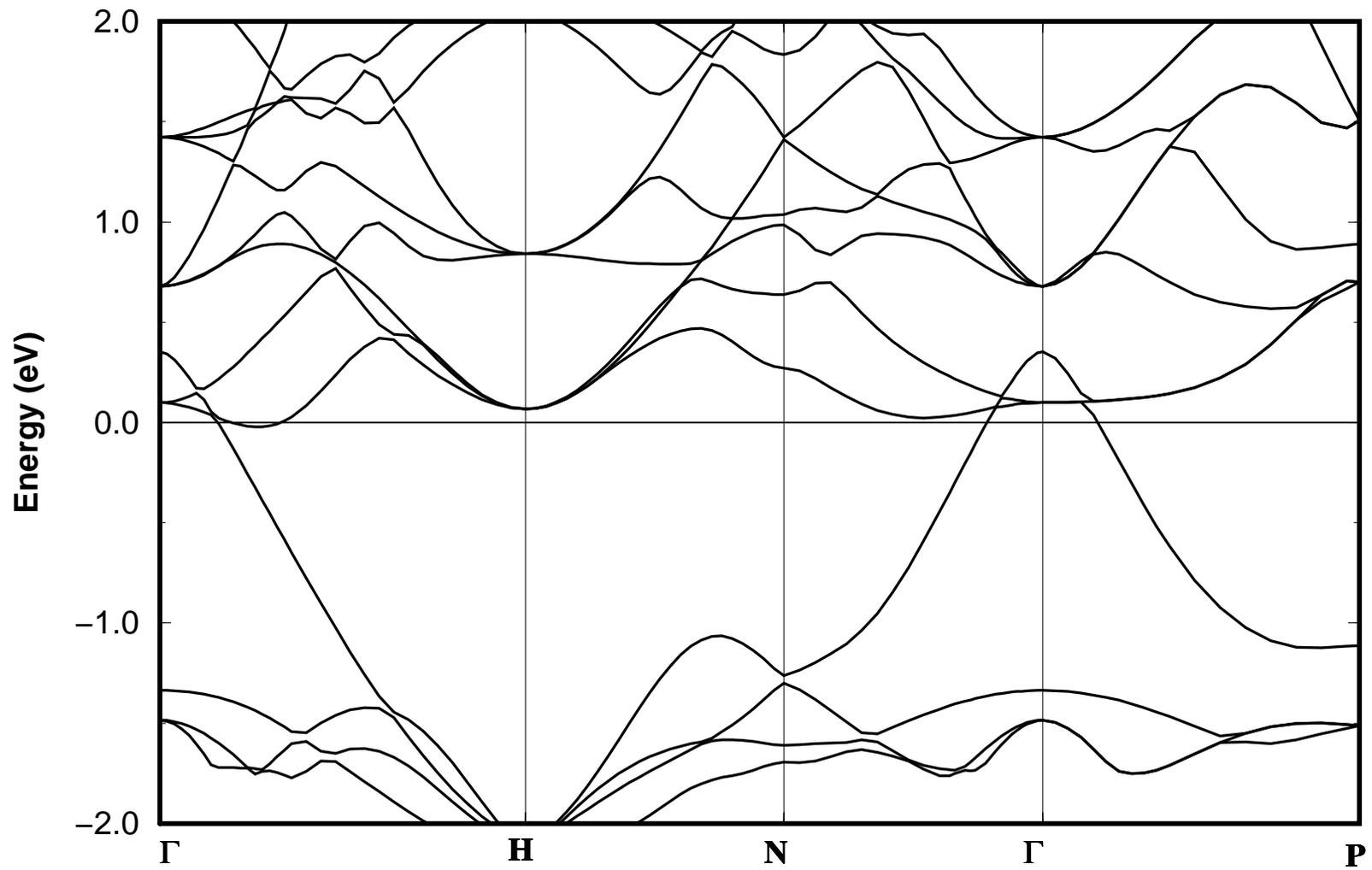
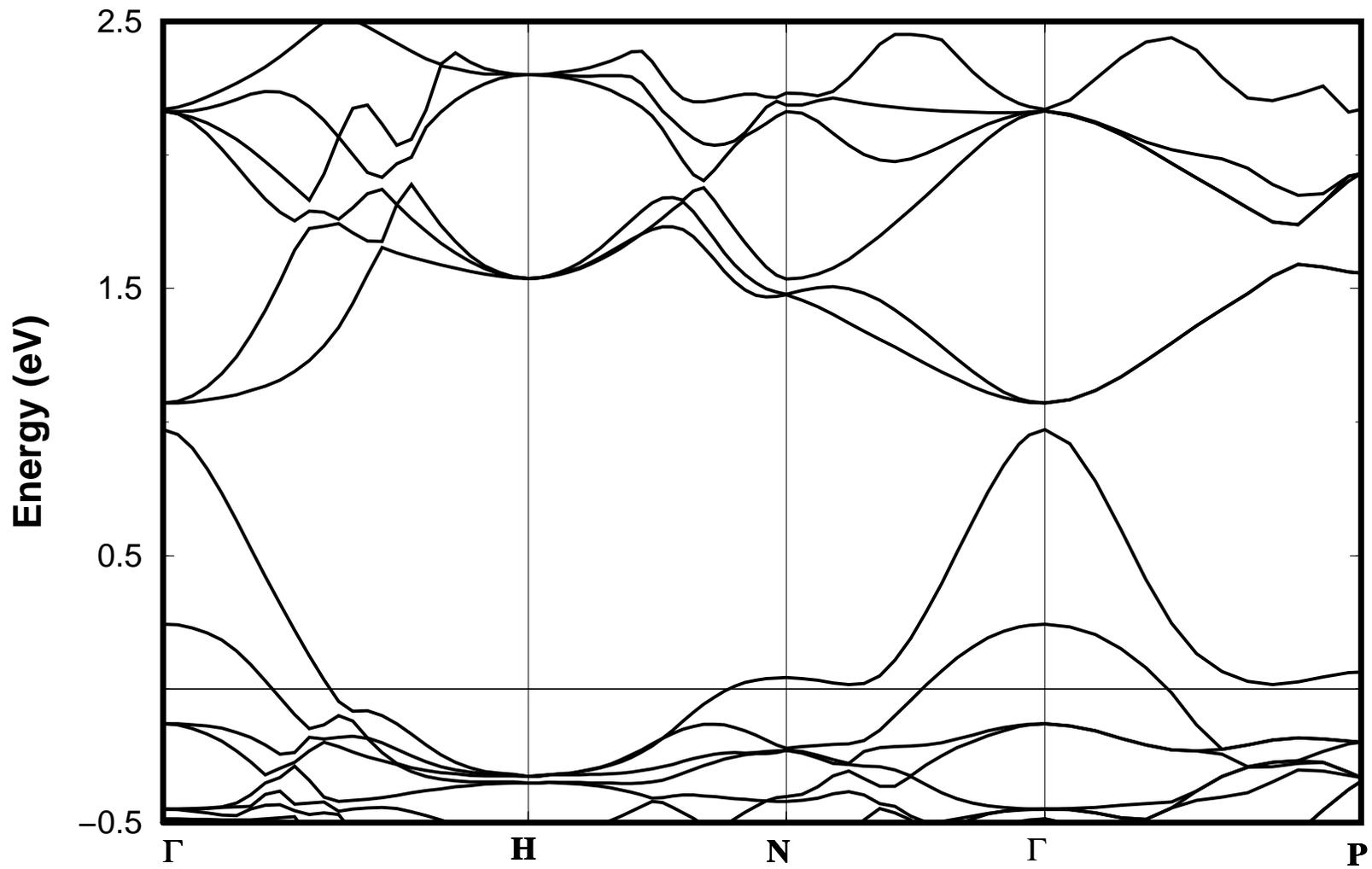
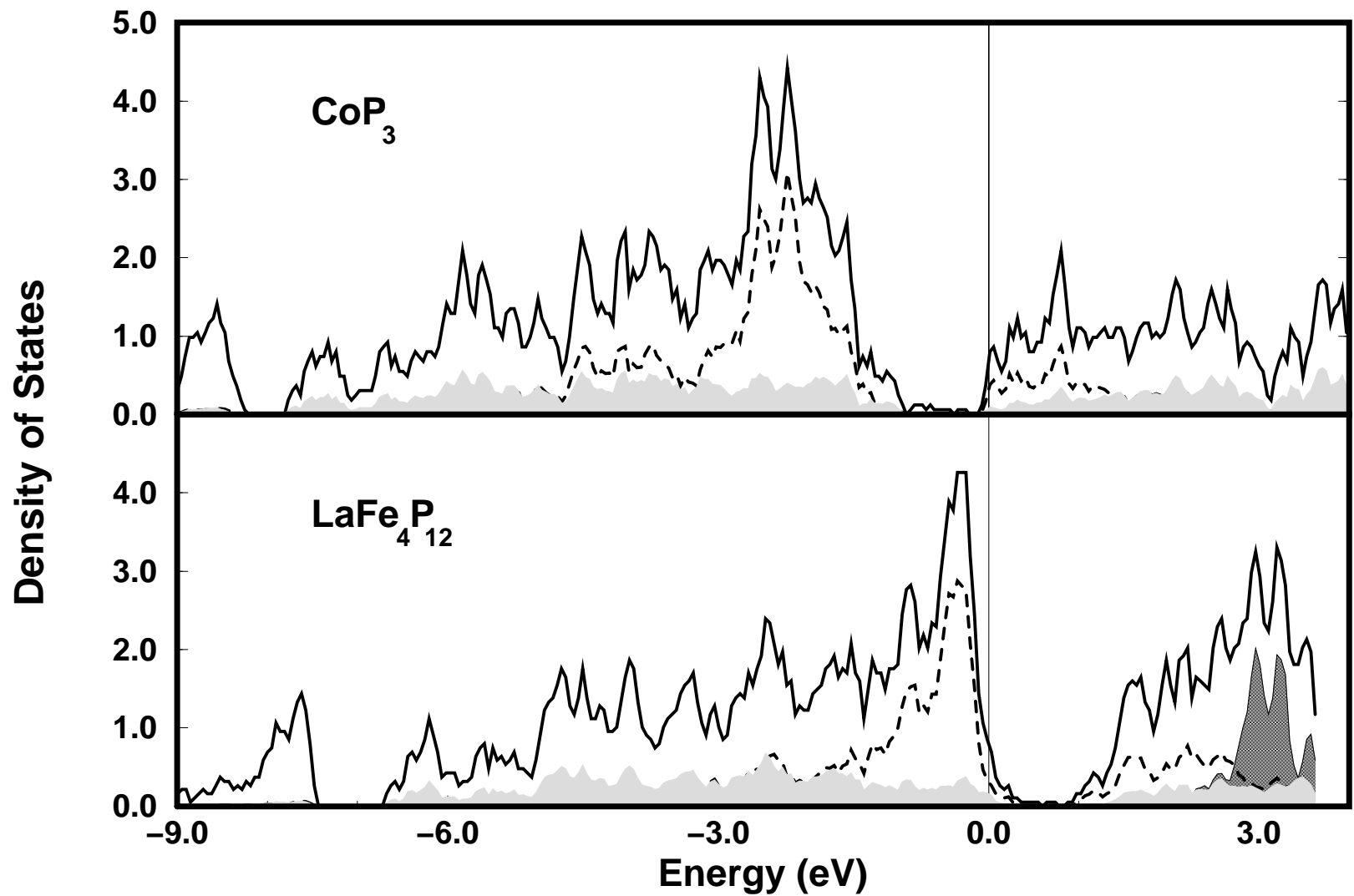
\end{document}